\documentclass[preprint,aps,amssymb,showpacs,superscriptaddress,nofootinbib]{revtex4}
\usepackage{graphicx}

\newcommand{\Case}[2]{{\textstyle \frac{#1}{#2}}}
\newcommand{\lP}{\ell_{\mathrm P}}

\newcommand{\sgn}{\mathop{\mathrm{sgn}}}

\begin{document}
%
\preprint{IMSc/2004/11/37}
\preprint{AEI-2004-100}

\title{Primordial Density Perturbation in
 Effective Loop Quantum Cosmology}

\author{Golam Mortuza Hossain}
\email{golam@imsc.res.in}
\affiliation{The Institute of Mathematical Sciences,
CIT Campus, Chennai-600 113, India.}
\affiliation{Max-Planck-Institut f\"ur Gravitationsphysik,
Albert-Einstein-Institut, Am M\"uhlenberg 1, D-14476 Potsdam,
Germany.}

\begin{abstract}
It is widely believed that quantum field fluctuation in an {\em
inflating} background creates the primeval {\em seed}
perturbation which through subsequent evolution leads to the
observed large scale structure of the universe. The standard
inflationary scenario produces {\em scale invariant} power
spectrum quite generically but it produces, unless {\em fine
tuned}, too large amplitude for the primordial density
perturbation than observed. Using similar techniques it is shown
that loop quantum cosmology induced inflationary scenario can
produce {\em scale invariant} power spectrum as well as {\em
small amplitude} for the primordial density perturbation without
fine tuning.  Further its power spectrum has a qualitatively
distinct feature which is in principle {\em falsifiable} by
observation and can distinguish it from the standard inflationary
scenario.
\end{abstract}

\pacs{04.60.Pp, 04.60.Kz, 98.80.Jk}

\maketitle


\section{Introduction}

The homogeneous and isotropic solution of general theory of
relativity, namely the Friedmann-Robertson-Walker (FRW) solution
appears to be an extremely good description of large scale
spacetime dynamics of our universe. Extreme simplicity of the FRW
solution nevertheless ignores some crucial features 
like it has certain sub-structure as well. On large
scale the deviation from homogeneity and isotropy being small
one can treat them as small perturbations around homogeneous
and isotropic background. The classical theory of large
scale structure formation in principle can be used to `derive' the
observed structure of current universe but these models need to
know the initial {\em seed} perturbations.  In this sense the
classical description of our universe is incomplete as there is
no mechanism of {\em generating} the seed perturbation within the
theory itself.

On the other hand the quantum field fluctuation in an {\em
inflating} background quite generically produces density
perturbation with scale-invariant power spectrum
\cite{Bardeen:DP} which is consistent with current observations.
This is certainly an attractive feature of the standard
inflationary scenario.  However, one major problem that plagues
almost all potential driven inflationary scenario that these
models generically produce too large amplitude for density
perturbation, typically $\frac{\delta \rho}{\rho} \sim 1-10^2$ at
horizon {\em re-entry} \cite{Brandenberger, Narlikar}. The cosmic
microwave background (CMB) anisotropy measurements on the other
hand indicates $\frac{\delta \rho}{\rho} \sim 10^{-5}$.
Naturally to make these models viable it is necessary to {\em
fine tune} the parameters of the field potential \cite{Guth}. In
the presence of quantum fluctuations it is rather difficult to
justify or sustain those fine tuning of field theoretical
parameters.

It is worthwhile to mention that inflation was invented to solve
some crucial problems of the Standard Big Bang (SBB) cosmology.
The most important of them is the so called {\em particle
horizon} problem. The horizon problem is directly related with
the fact the standard model of cosmology contains an initial
singularity where physical quantities like energy density,
spacetime curvature blow up leading to a breakdown of classical
description.  The initial singularity, however is viewed as an
attempt to extrapolate the classical theory beyond its natural
domain of validity. Near the classical singularity one expects
the evolution of the universe to be governed by a quantum
theory of gravity rather than the classical one.  Unfortunately
we are yet to formulate a completely satisfactory quantum theory
of gravity.

In recent years the issues regarding singularities in
cosmological models have been addressed in a rigorous way within
the framework of {\em loop quantum cosmology} (LQC)
\cite{cosmoI,cosmoII,cosmoIII,cosmoIV,IsoCosmo,SemiClass,LoopCosRev,Bohr}.
The loop quantum cosmology is a quantization of the cosmological
models along the line of a bigger theory known as {\em loop
quantum gravity} (LQG)
\cite{RovelliRev,ThiemannRevI,ThiemannRevII,AshtekarRev}. It has
been shown that the loop quantum cosmology cures the problem of
classical singularities in isotropic model \cite{Sing} as well as
less symmetric homogeneous model \cite{HomoLQC} along with
quantum suppression of classical chaotic behaviour near
singularities in Bianchi-IX models \cite{ChaosSup, BianchiIX}.
Further, it has been shown that non-perturbative modification of
the scalar matter Hamiltonian leads to a generic phase of
inflation \cite{Inflation, GenericInflation}. These features
crucially depends on a fact that the inverse scale operator
\cite{InvScale} in loop quantum cosmology has a bounded spectrum.
This is in a great contrast with classical situation where
inverse scale factor blows up as scale factor goes to zero.
However, not being a basic operator quantization of the inverse
scale factor operator involves quantization ambiguities
\cite{Ambig}.  One such ambiguity parameter referred as $j$, is
related with the dimension of representation of holonomy
operator, can take any half-integer value ({\em i.e.} $j \ge
\frac{1}{2}$).  The ambiguity parameter $j$ effectively controls
the duration of LQC induced inflationary phase. Being an
ambiguity parameter there is no unique way to fix the value of
$j$ within the loop quantum cosmology itself.

We have mentioned earlier that density perturbation generated by
quantum field fluctuation in an inflating background are believed
to be the {\em seed} perturbations responsible for the current
large scale structure of the universe. We have also mentioned
that non-perturbative modification of matter Hamiltonian in loop
quantum cosmology leads to a generic phase of inflation.
Naturally it is an important question to ask whether the density
perturbation generated by quantum fluctuations during loop
quantum cosmology induced inflationary phase can satisfy the
basic requirements of viability like scale-invariant power
spectrum. Further, it may leave some distinct imprint on the
power spectrum which may be observationally detectable as well.

Being inhomogeneous in nature treatment of these density
perturbation requires {\em inhomogeneous} models of loop quantum
cosmology. However the technology required to deal with
inhomogeneity at fundamental level within loop quantum cosmology
is {\em not} available yet.  Not having such technology, one
needs to proceed rather intuitively. Let us recall that in the
standard inflationary scenario for computing power spectrum of
density perturbation due to quantum fluctuations, one uses the
techniques which broadly can be classified as {\em Quantum Field
Theory in Curved Background} \cite{BirrellDavies, Ford}. In this
approach one treats the background geometry as classical object
whereas matter fields living in it are treated as quantum
entities. The main justification for using such techniques comes
from the fact the energy scale associated with inflationary
scenario is few order of magnitude lower than the Planck scale.
So one expects the geometry to behave more or less classically in
this regime.

In loop quantum cosmology in principle one can think of using
physical observables and physical inner product to evaluate the
physical expectation values to find out the behaviour of at least
the {\em homogeneous} part of the geometrical quantities.
Unfortunately developments of physical observables, physical
inner product and `time' evolution in loop quantum cosmology are
still in infant stage \cite{cosmoIV, HossainHO, Bojowald:Time}.
Nevertheless, one can construct an {\em effective} but {\em
classical} description of loop quantum cosmology using WKB
techniques. The effective loop quantum cosmology
\cite{EffectiveHam} incorporates important non-perturbative
modifications and has been shown to be generically non-singular
as well \cite{Singh:Bounce,Vereshchagin,GenBounce}.

In the effective loop quantum cosmology it has been shown
\cite{EffectiveHam} that in the region of interest (exponential
inflationary phase) gravitational part of Hamiltonian constraint
becomes same as the classical one with small quantum corrections.
However the scalar matter part of the effective Hamiltonian
remains non-perturbatively modified during this phase. In fact
non-perturbative modification of scalar matter Hamiltonian is
what that drives inflation in loop quantum cosmology. Having a
modified scalar matter Hamiltonian the scalar field satisfies a
{\em modified} Klein-Gordon equation instead of standard
Klein-Gordon equation. Naturally the mode functions of the scalar
field which contain the necessary information about background
geometry evolution and are essential to compute power spectrum of
the density perturbations, are expected to be different from the
standard mode-functions. Thus, although it may be justified to
employ similar techniques to compute the power spectrum in
effective loop quantum cosmology but certainly one {\em cannot}
borrow the same mode functions used in the standard inflationary
scenario.

In this paper we will compute power spectrum of density
perturbation using the {\em direct method} \cite{PaddyII}. In
this method one directly uses operator expression of `time-time'
component of stress-energy tensor (which is classically energy
density) to compute two-point {\em density correlation function}
and then evaluates its Fourier transform to compute power
spectrum of density perturbation. In the standard inflationary
scenario one generally avoids this direct computation as the two
point density correlation function in a pure classical background
diverges badly for small {\em coordinate length} separation ({\em
i.e.  ultra-violet divergence}). There usually one first computes
the power spectrum of field fluctuation. Using this one {\em
reconstructs} inhomogeneous but {\em classical} field
configuration which is then used to compute corresponding density
perturbation. However it is important to understand that this
divergence is rather {\em unphysical} because it arises when one
tries to resolve any two spatial points with arbitrary precision. 

In a quantum system, the expectation value of an operator which
is classically a phase space function in general is not equal to
the same function of the expectation values of basic phase space
operators. Thus the use of direct method is preferable over the
standard method as observational aspects deals with energy density
directly rather than the field configuration. Further, in the
standard method to relate the power spectrum of field
fluctuations with that of density perturbation, one needs to
know the {\em general} expression of the stress-energy tensor. In
this context, it is not yet a settled issue; how to obtain an
effective action from a quantum theory of gravity based on
canonical quantization.

In the context of standard inflationary scenario it was outlined
and explicitly shown \cite{PaddyII,PaddyI} that one can in fact 
regularize this field theoretical divergence by using the notion
of {\em zero-point} proper length.  Although it was used as an
{\em ad-hoc} assertion but it was argued that the notion of {\em
zero-point} proper length is expected from a proper theory of
quantum gravity. The power spectrums of density perturbations
computed using these two different method in the {\em relevant
energy scale} however are {\em not} very different. Nevertheless
there one can avoid rather cumbersome {\em indirect method} of
computing power spectrum of density perturbation.

In effective loop quantum cosmology, it has been shown that the
universe exhibits a generic {\em Big Bounce} with a non-zero
minimum {\em proper} volume \cite{GenBounce}. This in turns
implies a {\em zero-point} proper length for the isotropic
spacetime. Since the regularization technique is {\em naturally}
available in the effective loop quantum cosmology scenario then
it is quite appealing to directly use the operator expression of
`time-time' component of the stress-energy tensor to obtain the
power spectrum of the density perturbation due to the quantum
field fluctuation. In this sense this exercise can also be seen
as an explicit example of quantum gravity motivated
regularization technique to cure the {\em ultra-violet}
divergence of standard quantum field theory \cite{Thiemann:QSDV}.

In section II and III, we briefly review the standard scenario of
quantum field living in a DeSitter background and then obtain
corresponding two point density correlation function. In the next
section we review the scenario of effective loop quantum
cosmology which provides basic infrastructure required to
describe its inflationary phase. In particular we discuss about
the properties of the {\em effective equation of state} for the
scalar matter field. The effective equation of state essentially
summarizes the evolution of the background geometry. In the next
section we derive the modified Klein-Gordon equation which leads
to a modified mode function equation. We obtain an analytic
solution for the mode function equation. This modified mode
function reduces to the standard mode function in the appropriate
limit. Using the mode functions in the next section we compute
the power spectrum of the density perturbation. We discuss about
the properties of the power spectrum and its observational
implications.

\section{Quantum field in a De-Sitter Background}

In computing power spectrum of density perturbation in standard
inflationary scenario, one considers background geometry to be
homogeneous and isotropic. The invariant distance element in such
spacetime (using {\em natural units} {\em i.e.} $c=\hbar=1$) is
given by famous Friedmann-Robertson-Walker metric
\begin{equation}
ds^2 ~=~ -dt^2 ~+~ a^2(t)~d{\bf x}^2 ~,
\label{FRWMetric}
\end{equation}
where $a(t)$ is the {\em scale factor}. During inflationary
period the scale factor grows almost exponentially with
coordinate time. The Hubble parameter defined as $H:=\frac{\dot
a}{a}$ remains almost {\em constant} during the period. For
simplicity, in the intermediate period of calculation one treats
Hubble parameter as constant {\em i.e.} the evolution of
background geometry is considered to be De-Sitter like. One can
approximately compute the effect of small variation of Hubble
parameter on power spectrum, simply by considering the variation
of the final expression of power spectrum alone. This is in fact
a good approximation as the variation of Hubble parameter is
rather very small.

In standard scenario, the {\em inflation} is driven by scalar
field known as {\em inflaton} field. We will consider here the
most simple {\em single-field} inflationary scenario. The
dynamics of the scalar field is governed by the action 
\begin{equation} 
S_{\phi} ~=~ \int d^4 x \sqrt{-g}~ \left[ -\frac{1}{2} g^{\mu\nu}
\partial_{\mu}\phi \partial_{\nu}\phi - V(\phi) \right]
~=~ \int d^4 x \sqrt{-g} \mathcal{L}  ~.
\label{ScalarAction}
\end{equation} 
We have mentioned earlier that we will be using the direct method
to compute the power spectrum. So it will be quite useful to have
the expression of the stress-energy tensor for the scalar field.
The stress energy tensor corresponding to the action
(\ref{ScalarAction}) is given by 
\begin{equation} 
T_{\mu\nu} ~:=~ -~ \frac{2}{\sqrt{-g}} 
\frac{\delta S_{\phi}} {\delta g^{\mu\nu}} ~=~ 
\partial_{\mu}\phi \partial_{\nu}\phi~+~g_{\mu\nu} \mathcal{L} ~.
\label{StressTensor}
\end{equation} 
Comparing with the {\em perfect fluid} ansatz {\em i.e.} $
T_{\mu\nu} ~=~ (\rho + P )~ u_{\mu} u_{\nu}  ~+~ g_{\mu\nu} P$,
it is easy to see that $T_{00}$ component represents energy
density for the scalar field. In the canonical quantization one
treats Hamiltonian as a basic object.  Thus it is important for
the purpose of this paper to have the expression for the matter
Hamiltonian 
\begin{equation} 
H_{\phi} = \int d^3 x \left[ \frac{1}{2} a^{-3}  {\pi}_{\phi}^2 
~+~ \frac{1}{2} a {(\nabla \phi)}^2 
~+~ a^3 V(\phi) \right] ~,  
\label{SFIHamiltonian}
\end{equation} 
where ${\pi}_{\phi} = a^3 {\dot \phi}$. In deriving expression
(\ref{SFIHamiltonian}) it is assumed that the background {\em
geometry} is homogeneous and isotropic but {\em not} the scalar
field itself. This {\em approximation} can be justified as long
as the deviation from the homogeneity and isotropy remains small.
To make it more clear we rewrite the scalar field Hamiltonian
(\ref{SFIHamiltonian}) as
\begin{equation} 
H_{\phi} = a^{-3}~\int d^3 x \left[ \frac{1}{2} {\pi}_{\phi}^2
\right] 
~+~ a ~\int d^3 x \left[\frac{1}{2} {(\nabla \phi)}^2 \right]
~+~ a^3~\int d^3 x \left[ V(\phi) \right] ~.  
\label{SFIHamiltonian2}
\end{equation} 
In loop quantum cosmology, the geometrical quantities like the
scale factor $a$ here are represented through corresponding
quantum operators. While deriving effective classical Hamiltonian
from loop quantum cosmology, these operator expression
effectively get replaced by their corresponding eigenvalues.
The kinetic term of the scalar matter Hamiltonian
(\ref{SFIHamiltonian2}) involves inverse powers of the scale
factor. In loop quantum cosmology the inverse scale factor
operator has a bounded spectrum. Clearly one can see that the
kinetic term of the effective scalar matter Hamiltonian will
involve non-perturbative modifications due to loop quantization.
Using the Hamilton's equations of motion for the scalar field
{\em i.e.}
\begin{equation}
\dot \phi ~=~ \frac{\delta H_{\phi}}{\delta {\pi}_{\phi}}
~~~;~~~  \dot {\pi}_{\phi} ~=~ 
-\frac{\delta H_{\phi}}{\delta \phi} ~,
\label{HamiltonSFEOM}
\end{equation}
one can derive the second order equation of motion for the 
scalar field, given by
\begin{equation} 
\ddot \phi ~+~ 3 \left(\frac{\dot a}{a}\right) \dot \phi 
- \frac{\nabla^2 \phi}{a^2} + V'(\phi) ~=~ 0 ~.
\label{KleinGordonEquation}
\end{equation} 
The equation of motion (\ref{KleinGordonEquation}) for the scalar
field is the standard Klein-Gordon equation. It is worthwhile to
emphasize that one could have obtained the standard Klein-Gordon
equation (\ref{KleinGordonEquation}) simply by considering the
variation of the scalar field action (\ref{ScalarAction}). But
one should remember that our ultimate aim is to compute power
spectrum in effective loop quantum cosmology where
non-perturbative modification in the matter sectors comes through
its Hamiltonian.

To quantize the scalar field one proceeds in the standard way
{\em i.e.} by decomposing scalar field operator in terms of 
annihilation and creation operators $\hat{a}$ and $\hat{a^{\dagger}}$
as follows
\begin{equation} 
\hat{\phi}({\bf x}, t) ~=~ \int \frac{d^3 {\bf k}}{{(2\pi)}^3}
\left[ \hat{a}_k~ f_k(t)~ e^{i {\bf k \cdot x}} ~+~ 
\hat{a^{\dagger}}_k ~ f^*_k(t)~ e^{-i{\bf k \cdot x}} \right] ~,
\label{QuantizedSF}
\end{equation}
where $f_k(t)$ are the `properly normalized' mode functions.
Although one can quantize the scalar field analogous to that in
Minkowski spacetime, one faces the well known problem of defining
{\em vacuum state} in curved spacetime. In general, for curved
background geometry there does not exist an unique choice for the
vacuum state. Thus one needs to have some additional 
prescription to define it.

In the standard inflationary scenario one generally chooses so
called Bunch-Davies vacuum. It is defined as the state which gets
annihilated by $\hat{a}$ where the mode functions $f_k$ are so
`normalized' such that in `Minkowskian limit' {\em i.e.}, $H
\rightarrow 0$, the mode-function reduces to the flat space
positive frequency mode function $\frac{1}{\sqrt {2\omega}}~
e^{-i\omega t}$. We will use analogous definition for the vacuum
state for the calculation of power spectrum in effective loop
quantum cosmology as well. For simplicity we consider the
situation where the field potential is made of only the mass term
({\em i.e.} $V(\phi)=\frac{1}{2} m_{\phi}^2 \phi^2$). Then the mode
functions are the solution of the equation 
\begin{equation} 
\ddot f_k ~+~ 3 \left(\frac{\dot a}{a}\right) 
\dot f_k ~+~ \left(\frac{k^2}{a^2} + m_{\phi}^2\right)
f_k ~=~ 0 ~.
\label{MFEquation}
\end{equation} 
The mode function equation (\ref{MFEquation}) follows from the
Klein-Gordon equation (\ref{KleinGordonEquation}) and the
expansion of the scalar field operator (\ref{QuantizedSF}).  For
simplicity we consider the situation where the mass term can be
neglected ($\frac{k}{a} >> m_{\phi}$ ) in the equation
(\ref{MFEquation}). The `normalized' mode-function solutions
are then given by
\begin{equation} 
f_k ~=~ \frac{H}{\sqrt{2 k^3}} \left( 1 - i \frac{k}{H a}
\right)  e^{i\frac{k}{H a}} ~. 
\label{SModeFunction}
\end{equation} 
The mode function (\ref{SModeFunction}) in the `Minkowskian
limit' {\em i.e.}  $H \rightarrow 0$ limit reduces (upto a
constant phase) to flat space positive frequency mode function
$\frac{1}{\sqrt {2\omega}} e^{-i\omega t}$. This defines the
vacuum state $|0\rangle$ as $\hat{a} |0\rangle = 0$.

To compute the power spectrum of density perturbation using {\em
indirect method}, one first computes the power spectrum of field
fluctuation {\em i.e.} $\mathcal{P}_{\phi}(k) :=
\frac{k^3}{2\pi^2} {|f_k|}^2$. It is easy to see from the
expression of the normalized mode function (\ref{SModeFunction})
that at the time of horizon crossing ($a(t) = \frac{k}{2\pi H}$)
the corresponding power spectrum is {\em scale invariant}. In
getting mode function solution (\ref{SModeFunction}), we have
ignored the mass term of the scalar field. For the mass
dominating case ($\frac{k}{a} << m_{\phi}$ ), the `normalized'
mode functions are $f_k = \frac{1}{\sqrt{2 m_{\phi}}}
a^{-\frac{3}{2}} e^{-i m_{\phi}
t\left(\sqrt{1-{(\frac{3H}{2m_{\phi}})}^2}\right)}$.  It can be
easily checked that for this case also the corresponding power
spectrum is scale invariant at the time of horizon crossing. It
is often argued that the scale invariance is mainly determined
by the fact that during inflationary period the Hubble horizon
$H^{-1}$ remains almost constant. The details of particular model
of inflation has rather small effect on this property of the
power spectrum. 

\section{Two point density correlation function}

Having specified the vacuum state one can proceed to evaluate
vacuum expectation value of the two point density correlation
function. Two point density correlation function can naturally
be defined as 
\begin{equation} 
C({\bf x + l}, {\bf x}, t) ~:=~ \langle 0|~ \hat{T}_0^0({\bf x + l},t)~
 \hat{T}_0^0({\bf x},t) ~|0 \rangle ~. 
\label{TwoPointDCDef}
\end{equation} 
Using the expression of the scalar field operator
(\ref{QuantizedSF}) and the expression of the stress-energy
tensor (\ref{StressTensor}) one can evaluate the two point
density correlation function in terms of the mode-function,
given by \cite{PaddyII} 
\begin{equation} 
C({\bf x + l}, {\bf x}, t)
~=~ \int \frac{d^3 p}{{(2\pi)}^3} \frac{d^3 q}{{(2\pi)}^3}
~e^{i{\bf(p+q)\cdot l}}~ 
{\left|\dot f_p\dot f_q~-~
\left(\frac{{\bf p \cdot q}}{a^2} - m_{\phi}^2\right)
f_p f_q\right|}^2 ~.
\label{TwoPointDC1}
\end{equation} 
In evaluating two point density correlation function
(\ref{TwoPointDC1}), one ignores a space independent (formally
divergent) term as it would have contributed only to the $k=0$
mode while taking Fourier transform. Having known the normalized
mode function solution $f_k$ (\ref{SModeFunction}) one can
explicitly calculate the two point density correlation function
(\ref{TwoPointDC1}) given by 
\cite{PaddyII}
\begin{equation} 
C(l',t) ~:=~ 
C({\bf x + l}, {\bf x}, t) ~=~
\frac{1}{4\pi^4} ~\left[\frac{2 H^2}{(a l')^6}
 +  \frac{12}{(a l')^8}\right]  ~,
\label{TwoPointDC1e}
\end{equation} 
where $l'=|\bf{l}|$.
The expression (\ref{TwoPointDC1e}) of two-point density
correlation function {\em expectedly} diverges near $l'=0$.
However, as shown in \cite{PaddyII}, one can regularize this
divergence using the notion of zero-point proper length. The
expression (\ref{TwoPointDC1e}) in `Minkowskian limit' {\em i.e.}
$H\rightarrow 0$ limit reduces to the flat-space two point
density correlation function.

\section{Effective Isotropic Loop Quantum Cosmology}

In isotropic loop quantum cosmology, the basic phase space
variables are Ashtekar connection $c$ and densitized triad $p$.
The geometrical property of the space is encoded in the
densitized triad $p$ whereas the dynamics is encoded in the
connection $c$. In loop quantum cosmology one redefines
densitized triad to absorb the fiducial coordinate volume
component. This makes the proper volume of the
universe (\ref{FRWMetric}) to be $\int d^3 x \sqrt{-g} ~=~ a^3
V_0 ~=~ p^{\frac{3}{2}}$ \cite{Bohr}.

In loop quantum cosmology, the development of physical
observables and physical inner product are still in infant stage.
Nevertheless one can derive via WKB method, an effective
classical Hamiltonian which incorporate most important
non-perturbative modifications. This allows one to study the
effects of quantum modification using available standard tools.
The effective Hamiltonian for {\em spatially flat} isotropic loop
quantum cosmology derived in \cite{EffectiveHam}, is given by
\begin{equation}
H^{\text{eff}} =  -~\frac{1}{\kappa}
\frac{B_+(p)}{4 p_0} K^2 + W_{qg} + H^{\text{eff}}_{\phi}
\label{EffHamiltonian}
\end{equation}
where $\kappa = 16\pi G$, $p_0 = \frac{1}{6}\gamma \lP^2 \mu_0$,
$\gamma$ is the Barbero-Immirzi parameter, $K$ is the extrinsic
curvature (conjugate variable of $p$), $A(p)=
|p+p_0|^{\frac{3}{2}} - |p-p_0|^{\frac{3}{2}}$, $B_+(p) =
A(p+4p_0)+ A(p-4p_0)$, $\lP^2:=\kappa \hbar$ and $ W_{qg} =
\left(\frac{\lP^4}{288 \kappa p_0^3}\right) \left\{B_+(p) - 2
A(p) \right\}$. $\mu_0$ here is viewed as a quantization
ambiguity parameter and it is a order one number \cite{Bohr,
AshtekarRev}.  Apart from the modifications of the gravitational
kinetic term and scalar matter kinetic term, the effective
Hamiltonian (\ref{EffHamiltonian}) differs from the classical
Hamiltonian by a non-trivial potential term referred to as {\em
quantum geometry potential} $W_{qg}$. For the purpose of this
paper we will be interested in the regime $(p_0<<p)$ where the
quantum geometry potential has natural interpretation of being
{\em perturbative} homogeneous quantum fluctuations around FRW
background. The effective scalar matter Hamiltonian is given by
\begin{equation}
H^{\text{eff}}_{\phi} =
\frac{1}{2} {|\tilde{F}_{j,l}(p)|}^{\frac{3}{2}} {p_{\phi}}^2
~+~ p^{\frac{3}{2}} V(\phi) ~,
\label{EffMatterHam}
\end{equation}
where $p_{\phi} (= V_0 \pi_{\phi})$ is the field {\em momentum},
$\tilde{F}_{j,l}(p)$ is the eigenvalue of the inverse densitized
triad operator $\hat{p^{-1}}$ and is given by $\tilde{F}_{j,l}(p)
= (p_j)^{-1} F_l( p/p_j )$ where $p_j =\frac{1}{3}\gamma\mu_0 j
l_p^2$. The $j$ and $l$ are two quantization ambiguity parameters
\cite{Ambig,ICGCAmbig}. The half integer  $j$ is related with the
dimension of representation while writing holonomy as
multiplicative operators. The real valued $l$ ($0<l<1$)
corresponds to different, classically equivalent ways of writing
the inverse power of the densitized triad in terms of Poisson
bracket of the basic variables. The function $F_l(q)$ is given by
\cite{BianchiIX}
\begin{eqnarray}
F_l(q)&:=& \left[ \frac{3}{2(l+2)(l+1)l} \left( ~
(l+1) \left\{ (q + 1)^{l+2} - \left. |q - 1|^{l+2} \right\} ~-~ 
\right. \right. \right. \nonumber \\
& & ~~~(l+2) q \left\{ (q + 1)^{l+1} - \right. 
 \left. \left. \left. \sgn(q - 1) |q - 1|^{l+1} \right\}
~ \right) ~\right]^{\frac{1}{1-l}} \nonumber \\
& \rightarrow &  q^{-1}  ~~~~~~~~~~~~~~~~~(q \gg 1) \nonumber \\
& \rightarrow &  \left[ \frac{3 q}{l+1}\right]^{\frac{1}{1-l}}
~~~~~~~(0 < q \ll 1) ~. 
\label{InvSF}
\end{eqnarray}
From the expression (\ref{InvSF}) one should note that for large
values of the densitized triad {\em i.e.} in large volume one has
the expected classical behaviour for the inverse densitized triad.
The quantum behaviour is manifested for smaller values of the
densitized triad. Here the meaning of large or small values of
the triad $p$ is determined necessarily by the values of
$p_j$. The quantum mechanically allowed values for the ambiguity
parameter $l$ is ($0<l<1)$. Now one should also note that if one
takes the ambiguity parameter value $l=2$ then the small volume
expression (\ref{InvSF}) becomes same as the large volume
expression. In other words, taking ambiguity parameter value
$l=2$ is equivalent of taking large volume limit {\em i.e.}
classical limit. This observation will be very useful in fixing
the choice of vacuum while computing two-point density
correlation function in this effective background.

In loop quantum cosmology $p_0$ and $p_j$ represent two important
(square of) length scale. $p_0$ demarcate the {\em strong}
quantum effect regime ({\em non-perturbative} regime) from {\em
weak} quantum effect regime ({\em perturbative} regime) of the
{\em gravity sector} whereas $p_j$ demarcate the {\em same} for
the {\em matter sector}. Since ambiguity parameter
$j\ge\frac{1}{2}$, so it follows from their respective definition
that $p_j \ge p_0$. Naturally, {\em non-perturbative modification
of matter sector can survive longer than the same for the gravity sector
} depending on the value of the ambiguity parameter
$j$. We have mentioned earlier that in computing power spectrum
of density perturbation we will use similar techniques used in
the standard inflationary scenario. In this approach one treats
geometry as a classical object whereas matter fields living in it
are treated as quantum objects. Thus self-consistency of this
framework requires that we should consider the regime where
$p>>p_0$ in our calculation. In this regime the gravitational
part of the Hamiltonian constraint becomes same as the classical
Hamiltonian with small quantum correction. The reduced effective
Hamiltonian in this regime is given by
\begin{equation}
H^{\text{eff}} =  -~\frac{3}{2\kappa}
K^2 {\sqrt p} ~-~\frac{\lP^4}{24\kappa} p^{-\frac{3}{2}} 
 ~+~ H^{\text{eff}}_{\phi} ~.
\label{EffHamLarge}
\end{equation}
The loop quantum cosmology induced inflationary scenario persist
as long as densitized triad $p$ remains less than $p_j$. Thus we
will be interested in computing power spectrum of density
perturbation in the regime ($p_0<<p <p_j$).  In this regime the
{\em effective} energy density and pressure are given by
$\rho^{\text{eff}}= p^{-\frac{3}{2}}H^{\text{eff}}_{\phi} $ and
$P^{\text{eff}}= -\frac{1}{3}p^{-\frac{3}{2}} (2 p~\frac{\partial
H^{\text{eff}}_{\phi}}{\partial p})$ \cite{EffectiveHam}.  It can be
checked easily using relation between scale factor and densitized
triad that these definition satisfy standard {\em conservation}
equation $a \frac{d \rho^{\text{eff}}}{d a} = -3
(\rho^{\text{eff}} + P^{\text{eff}})$. Furthermore one can recover
standard expression of energy density and pressure using the
standard scalar matter Hamiltonian $H_{\phi}$ in place of the modified
scalar matter Hamiltonian $H^{\text{eff}}_{\phi}$ in these definition.
It is shown in \cite{GenericInflation} that the effective
equation of state $\omega^{\text{eff}}:=
P^{\text{eff}}/{\rho^{\text{eff}}}$ can be expressed as a
function of standard equation of state $\omega$ and the
densitized triad $p$
\begin{equation}
\omega^{\text{eff}} =  -1 + \frac{(1+\omega)p^{\frac{3}{2}}
[\tilde{F}_{j,l}(p)]^{\Case{3}{2}} \left(1 -
\frac{p}{\tilde{F}_{j,l}(p)} \frac{d \tilde{F}_{j,l}(p)}{d p}
\right)}
{(1+\omega)p^{\frac{3}{2}}[\tilde{F}_{j,l}(p)]^{\frac{3}{2}}  +
(1-\omega)} ~.	
\label{EffectiveEOS}
\end{equation}

Using the expression (\ref{InvSF}) it is easy to see that for the
large values of the densitized triad $p$, where one expects the
quantum effects to be small, $\omega^{\text{eff}} = \omega$
whereas for small values of $p$ the $\omega^{\text{eff}}$ differs
from the classical $\omega$ dramatically. In this paper we will
be interested in the situation where $\omega^{\text{eff}}\approx
-1$ (for $p<p_j$). This requirement will automatically be
satisfied if at the end of loop quantum cosmology induced
inflation the radiation or matter domination or even another
phase of classical acceleration ({\em i.e} $\omega = \frac{1}{3},
0, <-\frac{1}{3}$) begins. Thus, during the loop quantum
cosmology induced inflationary period one can express the matter
Hamiltonian as
\begin{equation}
H^{\text{eff}}_{\phi} \approx  \bar{\rho} ~p^{\frac{3}{2}} ~, 
\label{MatterHam}
\end{equation}
where $\bar{\rho}$ is a constant of integration. Physically
$\bar{\rho}$ corresponds to the {\em maximum} energy density that
can be `stored' in the effective spacetime. This also defines the
energy scale associated with the loop quantum cosmology induced
inflationary scenario. 

It has been shown in \cite{GenBounce} that the effective loop
quantum cosmology exhibits a generic bounce with non-zero minimum
{\em proper} volume. It follows from the equation
(\ref{EffHamLarge}) and the equation (\ref{MatterHam}) that the
minimum value of the proper distance $L_0$, defined as $L_0^2 :=
p_{min} = p( H^{\text{eff}}=0; K=0) $, is given by
\begin{equation}
L_0^6  ~=~ \frac{2\pi G}{3~ \bar{\rho}} ~.
\label{ZeroPointPL}
\end{equation}
Self-consistency of the expression (\ref{ZeroPointPL})
requires $p_0<<p_{min}<p_j$.

In standard cosmology one uses the scale factor as geometric
variable. In isotropic loop quantum cosmology, the basic variable
is densitized triad $p$ defined as $p^{\frac{3}{2}} := \int d^3 x
\sqrt{-g} = a^3 V_0$, where $V_0$ is fiducial {\em coordinate}
volume.  Clearly the densitized triad $p$ here is a dimensionful
quantity whereas the scale factor $a$ is dimensionless. Also the
absolute value of the scale factor is physically irrelevant.
Rather what matters is the ratio of scale factor at two different
period. Naturally there is a freedom left in relating the scale
factor with the densitized triad. We define the relation between
the scale factor and the eigenvalues of densitized triad operator
such that for small volume limit 
\begin{equation}
\hat{p} ~ \hat{p^{-1}}~ |\mu \rangle  ~:=~
 a^{2(1+ \frac{1}{1-l})} ~|\mu \rangle ~. 
\label{SFDefinition}
\end{equation}
We have mentioned earlier that taking ambiguity parameter value
$l=2$ is equivalent of taking large volume limit of the inverse
densitized triad spectrum. Clearly in our choice of definition
the scale factor takes the value $a=1$ at the transition point
from non-perturbatively modified matter sector to the standard
matter sector. For the regime $p<p_j$ one can approximate the
effective equation of state (\ref{EffectiveEOS}) as
\begin{equation}
(1+\omega^{\text{eff}})~\simeq~ C_{\omega}~
\left(\frac{n+2}{3}\right) ~a^{2(1-n)} ~,
\label{EffEOS}
\end{equation}
where $C_{\omega} = 2\left(\frac{1+\omega}{1-\omega}\right)$ and
$n = -\frac{1}{2}(1+ \frac{3}{1-l})$.  The last two terms in the
effective Hamiltonian constraint (\ref{EffHamLarge}) are
comparable near bounce point. However once the densitized triad
$p$ starts increasing then it is clear from the equation
(\ref{EffHamLarge}) that the contribution from quantum geometry
potential quickly drops out compared to the matter Hamiltonian
(\ref{MatterHam}). Naturally for the region away from the bounce
point one can write down the Hamiltonian constraint
($H^{\text{eff}}=0$) in terms of the scale factor as
\begin{equation}
3~{\left(\frac{\dot a}{a}\right)}^2 ~\simeq~ 8\pi G~ \bar{\rho} ~,
\label{FriedmannEqn}
\end{equation}
where we have used the Hamilton's equation of motion 
${\dot p} = \frac{\kappa}{3} \frac{\partial H^{eff}}{\partial K}$.
The equation (\ref{FriedmannEqn}) is nothing but the usual
Friedmann equation.  Using the equation (\ref{ZeroPointPL}) and
the equation (\ref{FriedmannEqn}) we can define a dimensionless
quantity
\begin{equation}
\sigma ~:=~ 2\pi H L_0 ~=~ 4\pi 
{\left(\frac{2\pi}{3}\right)}^{\frac{2}{3}}
~{\left(\frac{\bar{\rho}}{M_p^4}\right)}^{\frac{1}{3}} ~,
\label{CutoffScale}
\end{equation}
where $G=M_p^{-2}$. This will be a useful quantity in calculation
of power spectrum. Since we consider the situation where
$p_0<<L_0^2<p_j$ {\em i.e.} bounce occurs at a time when proper
volume of the universe is much larger than the Planck volume.
Thus it is clear that $\sigma$ is much smaller than unity
($\sigma<<1$) during the loop quantum cosmology induced
exponential inflationary phase.

From the definition of FRW metric (\ref{FRWMetric}) it follows
that the {\em proper} distance square say $d^2(a,l')$, between two
points separated by {\em coordinate distance} $l'$ on a given
spatial slice ($dt=0$) is simply $d^2(a,l') = (a~ l')^2$. In other
word, in the classical geometry the proper distance between two
points is simply `coordinate distance times the scale factor'.
In classical case one can choose coordinate distance separation
arbitrarily small. Naturally the proper distance between two
points can become arbitrarily small. In loop quantum cosmology
the basic variable is a densitized triad instead of the usual
metric variable.  Further in loop quantum cosmology, one
redefines the densitized triad by absorbing component of the
fiducial coordinate volume. This makes the proper volume of the
universe to be just $p^{\frac{3}{2}}$. In case of effective loop
quantum cosmology, we have seen that there exist a {\em non-zero}
minimum value for the densitized triad $p$. To incorporate such
feature in the definition of the proper distance in effective
loop quantum cosmology, we introduce the notion of {\em effective
coordinate length} $l^{\text{eff}}(a,l')$. The proper distance
between two points separated by coordinate distance $l'$ is defined
as
\begin{equation}
d^2(l',a) ~:=~ (a~ l^{\text{eff}})^2 ~=~ L_0^2 + (a~l')^2 ~.
\label{EffCoordinateL}
\end{equation}
The effective coordinate length keeps the usual notion of proper
distance {\em i.e.} `coordinate distance times scale factor'
intact and incorporate feature like zero-point proper length.
Further, it allows one to use the standard machinery while
computing the power spectrum of density perturbation and acts as
an ultra-violet regulator of standard quantum field theory. For
large volume ({\em i.e.} $(a l')$ large) this definition is
virtually equivalent to the standard definition of proper
distance as $L_0$ is very small (a few Planck units).

\section{Modified Klein-Gordon Equation}

We have mentioned earlier that the kinetic term of the scalar
matter Hamiltonian gets non-perturbative modification as its
classical expression involve inverse powers of densitized triad.
The effective scalar matter Hamiltonian obtained as outlined in
the previous section is given by
\begin{equation} 
H^{\text{eff}}_{\phi} = V_0 {|\tilde{F}_{j,l}(p)|}^{\frac{3}{2}} 
\int d^3 x \left[ \frac{1}{2} {\pi}_{\phi}^2 \right] 
~+~ V_0^{-\frac{1}{3}} p^{\frac{1}{2}} \int d^3 x 
\left[\frac{1}{2} {(\nabla \phi)}^2 \right]
~+~ V_0^{-1} p^{3/2} \int d^3 x \left[ V(\phi) \right] ~.  
\label{EffectiveHamiltonian}
\end{equation} 
It should be noted that we have now kept the gradient term in the
effective Hamiltonian. Earlier while computing background
evolution the gradient term was neglected as one assumes that the
background evolution is mainly determined by the homogeneous and
isotropic contribution of the matter Hamiltonian. In other words
the inhomogeneity is assumed to be small. Using the Hamilton's
equations of motion for the effective Hamiltonian
(\ref{EffectiveHamiltonian}) one can derive the corresponding
{\em modified} Klein-Gordon equation, given by
\begin{equation} 
\ddot \phi ~-~ 3 \left(\frac{1}{1-l} \right)
\left(\frac{\dot a}{a}\right)
\dot \phi ~+~ a^{3 + \frac{3}{1-l}}
 \left( - \frac{\nabla^2 \phi}{a^2} + V'(\phi) \right) ~=~ 0 ~,
\label{MKGEquation}
\end{equation} 
where we have substituted eigenvalue of the inverse triad
operator by scale factor using the definition (\ref{SFDefinition}).
It is easy to see that if one takes the value of the ambiguity
parameter $l=2$ then the modified Klein-Gordon equation 
(\ref{MKGEquation}) goes back to the standard Klein-Gordon
equation (\ref{KleinGordonEquation}). 

It is important to emphasize here that the non-perturbative
modification of the scalar matter Hamiltonian which is being
studied here, comes from the bounded spectrum of the inverse
scale factor operator. Since the modification affects the kinetic
term of the scalar matter Hamiltonian, it essentially affects all
the modes. It can be seen from the equation (\ref{MKGEquation})
as well. This modification is distinct from the other Planck
scale effects studied in the literature. For example, in the
context of trans-Planckian inflation
\cite{Brandenberger:TPI,Niemeyer}, one studies the possible
effects of Planck scale modification of the dispersion relation
or the possible effects of the space-time non-commutativity
\cite{Tsujikawa:NCI,Brandenberger:NCI}.

\section{Modified Mode Functions}

Using the expression for the quantized scalar field
(\ref{QuantizedSF}) like in standard case one can derive the {\em
modified} mode function equation for the scalar field
\begin{equation} 
\ddot f_k ~-~ 3 \left(\frac{1}{1-l} \right)
\left(\frac{\dot a}{a}\right) \dot f_k ~+~ a^{3 + \frac{3}{1-l}}
~H^2~\left(\frac{k^2}{H^2 a^2} + \frac{m_{\phi}^2}{H^2}\right)
f_k ~=~ 0 ~. 
\label{ModifiedMFEquation}
\end{equation} 
One can easily check that for $l=2$ ({\em i.e. the classical
case}) the modified mode function equation goes back to the
standard mode-function equation (\ref{MFEquation}). To compute
the power spectrum it is essential to know the solution of the
mode function equation (\ref{ModifiedMFEquation}). For simplicity
we will neglect the mass term {\em i.e.} we will assume
($\frac{k}{H a} >> \frac{m_{\phi}}{H}$).  To simplify the mode
function equation (\ref{ModifiedMFEquation}) further, we make
change of variables as follows 
\begin{equation}
f_k := a^{-n} \tilde{f_k} ~;~ dt = a^n d\eta 
\label{NewMFVariables}
\end{equation}
with the value of $n = -\frac{1}{2}(1+ \frac{3}{1-l})$.  In loop
quantum cosmology allowed values for the $l$ ambiguity parameter
is $(0<l<1)$ whereas the classical situation can be obtained
simply by taking $l=2$. In terms the new parameter $n$, the
classical situation corresponds to $n=1$ and quantum situation is
described for ($-\infty<n<-2$).  One may note here that for $n=1$
the new variable $\eta =-\frac{1}{n H a^n}$ is nothing but {\em
conformal} time.  In terms of these new variables
(\ref{NewMFVariables}) the mode function equation
(\ref{ModifiedMFEquation}) becomes 
\begin{equation}
\frac{d^2 \tilde{f_k}}{d\eta^2} ~-~ \left(2~\frac{1}{2n} -
1\right)\frac{1}{\eta} \frac{d \tilde{f_k}}{d\eta} ~+~ \left(k^2
+\frac{(\frac{1}{2n})^2 -(1+\frac{1}{2n})^2}{\eta^2}
\right)\tilde{f_k} ~=~ 0 ~.  
\label{NewModifiedMFEquation}
\end{equation} 
The equation (\ref{NewModifiedMFEquation}) is a modified
expression of Bessel differential equation  and admits analytical
solution of the form \cite{Bowman} 
\begin{equation}
\tilde{f_k} ~=~ \eta^{\frac{1}{2n}} \left[ A_{(k,n)}
J_{-(1+\frac{1}{2n})}(k\eta) +  B_{(k,n)}
J_{(1+\frac{1}{2n})}(k\eta) \right]  ~, 
\label{NewMFSolution}
\end{equation}
where $A_{(k,n)}$ and $B_{(k,n)}$ are two constants of
integration corresponding to {\em second} order differential
equation of the mode-function. 

To fix these constants of integration we require that for large
volumes ($n=1$) the modified mode function reduces to the
standard `normalized' mode function (\ref{SModeFunction}). Since
the standard mode function (\ref{SModeFunction}) are already
`normalized' to pick out the Bunch-Davies vacuum then this
requirement will automatically fix the choice of vacuum in
effective loop quantum cosmology. This fixes the mode function
solution to be 
\begin{equation}
f_k ~=~ {\sqrt \frac{n+2}{3}} (-n
H) {\sqrt \frac{\pi}{4 k^3}}~ (k \eta)^{1+\frac{1}{2n}} \left[
J_{-(1+\frac{1}{2n})}(k\eta) ~+~  i~ J_{(1+\frac{1}{2n})}(k\eta)
\right]  ~.
\label{NormalizedMFSolution}
\end{equation}
Using Bessel function identities $J_{n+1}(x) + J_{n-1}(x) =
\frac{2n}{x} J_n(x)$,  $J_{-\frac{1}{2}} = \sqrt{\frac{2}{\pi
x}}~ cos(x)$ and $J_{\frac{1}{2}} = \sqrt{\frac{2}{\pi x}}~
sin(x)$ one can easily check that for $n=1$ the modified mode
function (\ref{NormalizedMFSolution}) reduces to the standard
mode function (\ref{SModeFunction}).

It is worth pointing out that in the expression
(\ref{NormalizedMFSolution}) we have specifically chosen the
power of $(\frac{n+2}{3})$ to be $\frac{1}{2}$. But it is clear
that for any arbitrary power of $(\frac{n+2}{3})$, the mode
function would reduce to the standard mode function. We have made
this choice precisely to absorb similar term coming from the
effective equation of state (\ref{EffEOS}) that appears in the
final expression of power spectrum. In other words we have chosen
the vacuum state such that it satisfies Bunch-Davies prescription
{\em and} the computed power spectrum is free from trivial {\em
ambiguity parameter dependent} multiplicative factor.

\section{Power Spectrum}

Having known the exact solution of the mode function in principle
one can evaluate the two-point density correlation function using
the expression (\ref{TwoPointDC1}). Let us recall that we are
mainly interested in finding out the power spectrum at the time
of horizon crossing {\em i.e.} $\frac{k}{H a} = 2 \pi$. The
argument of the Bessel function $k\eta = \frac{k}{H a}
\left(\frac{a^{1-n}}{-n}\right) << 1$ during horizon crossing.
For super-horizon scale above inequality holds naturally; even
for sub-horizon scale upto reasonable extent the same inequality
will hold since for effective loop quantum cosmology
($-\infty<n<-2$). Since the asymptotic form of the Bessel
function is $J_{m}(x) \approx \frac{1}{\Gamma(1+m)}
(\frac{x}{2})^m$ for $x<<1$ then clearly the dominating
contribution in the mode function (\ref{NormalizedMFSolution})
comes from the first term.  So we will approximate the mode
function as
\begin{equation} 
f_k ~\approx~ {\sqrt \frac{n+2}{3}} (-n H) {\sqrt \frac{\pi}{4 k^3}}~
(k \eta)^{1+\frac{1}{2n}}
\left[ J_{-(1+\frac{1}{2n})}(k\eta) \right]  ~,
\label{ApproxMF}
\end{equation} 
for further evaluation and will use its asymptotic form for
explicit calculation.

Using the expression of two-point density correlation function
(\ref{TwoPointDC1}) and expression of mode function 
(\ref{ApproxMF}) one can simply follow the similar steps as in
\cite{PaddyII} to derive the expression of two point density 
correlation function
\begin{equation} 
C(l',t) ~=~ 
{\left(\frac{n+2}{3}\right)}^2 \frac{a^{4(1-n)}}
{2^{2-\frac{2}{n}}~ {(2\pi)}^2~ {\Gamma(1-\frac{1}{2n})}^4 }
\left[ \frac{2 H^2}{(a l')^6} +  \frac{a^{4(1-n)}}{(a l')^8} \right] ~.
\label{TwoPointDC2}
\end{equation} 
In deriving the above expression (\ref{TwoPointDC2}), it is quite
helpful to use the Bessel function identity $\frac{d}{d x}[x^m
J_{-m}(x)] = - x^m J_{1-m}(x)$ while evaluating time derivative
of the mode function (\ref{ApproxMF}). Now it is easy to check
that for $n=1$ ({\em i.e.} classical mode function) the
expression becomes qualitatively same as (\ref{TwoPointDC1e}).
But we may note that it is {\em quantitatively} slightly
different than (\ref{TwoPointDC1e}). The source of this
difference can be traced back to the approximation that we have
made. In the case of loop quantum cosmology the argument of the
Bessel function $k\eta = \frac{k}{H a}
\left(\frac{a^{1-n}}{-n}\right) << 1$ at the time of horizon
crossing. But clearly the same does {\em not} hold for the
standard case ($n=1$).

The two point density correlation function (\ref{TwoPointDC2})
diverges as the ``coordinate length" $l'$ goes to zero. This
feature is rather expected from a calculation based on standard
quantum field theory. However as we have mentioned that one can
regularize this expression using the notion of zero-point proper
length which is naturally available in effective loop quantum
cosmology.  In section IV, we have introduced the notion of {\em
effective coordinate length}. This basically allows one to use
the available machinery used in the standard case.  Essentially
this step summarizes the ultra-violet regularization of two point
density correlation function.  We define {\em effective} two
point density correlation function as the {\em regularized} form
of the standard two point density correlation function as
\begin{equation} 
C^{eff}(l',t) ~:=~ C(l^{eff},t) ~. 
\label{EffTwoPointDC}
\end{equation} 

Now we can evaluate the Fourier transform of the effective
two point density correlation in usual way 
\begin{eqnarray} 
|\rho_{k}(t)|^2 ~&:=&~ \int ~d^3 {\bf l} ~e^{i {\bf k \cdot l}}~
C^{eff}(l',t) \nonumber \\
~ &=&~   
{\left(\frac{n+2}{3}\right)}^2 \frac{a^{4(1-n)}}
{2^{2-\frac{2}{n}}~ {(2\pi)}^2~ {\Gamma(1-\frac{1}{2n})}^4 }
\left[ \frac{2 H^2}{a^6}I_1 ~+~  \frac{a^{4(1-n)}}{a^8}~ I_2 \right] ~,
\label{DensityModeSquare}
\end{eqnarray} 
where the integrals $I_1$ and $I_2$ can be evaluated using method
of contour integration. They are given by
\begin{equation}
I_1 := 
\int \frac{d^3 {\bf l}~ e^{i {\bf k \cdot l}}}
{ {({l'}^2 + \frac{L_0^2}{a^2})}^{3}}
=  \frac{\pi^2 e^{-\frac{k L_0}{a}} }{4} 
{\left(\frac{a}{L_0}\right)}^3
\left[ 1 + \frac{k L_0}{a}\right] ~,
\label{CI1}
\end{equation}
and
\begin{equation}
I_2 := 
\int \frac{d^3 {\bf l}~ e^{i {\bf k \cdot l}}}
{ {({l'}^2 + \frac{L_0^2}{a^2})}^{4}}
= \frac{\pi^2 e^{-\frac{k L_0}{a}} }{8} {\left(\frac{a}{L_0}\right)}^5
\left[ 1 + \frac{k L_0}{a} + \frac{1}{3} 
{\left( \frac{k L_0}{a}\right)}^2 \right] ~. 
\label{CI2}
\end{equation}

The power spectrum of density perturbation generated during
inflation however is not directly observable. Rather the observed
power spectrum corresponds to the density perturbation at the
time of horizon {\em re-entry} in the post-inflationary period.
In the intermediate period between horizon exit and horizon
re-entry the density contrast $\delta(:=\frac{\delta
\rho}{\rho})$ remains almost constant for the super-horizon
modes. Nevertheless due to the change in equation of state of the
total matter field leads to a scaling of the amplitude of density
perturbation. Super-horizon evolution in Bardeen's gauge invariant
formalism \cite{Bardeen:GIP} of density perturbation leads to a
rather simple formula for the evolution of density contrast  
\begin{equation}
\left| \frac{\delta_k}{1+\omega}\right|_{t=t_f} \approx 
\left| \frac{\delta_k}{1+\omega}\right|_{t=t_i} ~,
\label{BardeenFormula}
\end{equation}
where $\delta_k := \frac{\rho_k(t)}{\bar{\rho}}$,  $t_i$ and
$t_f$ are initial and final time respectively.  The power
spectrum of density perturbation at the time of {\em re-entry} is
given by
\begin{equation}
\mathcal{P}_{\delta}(k) ~=~ \frac{k^3}{2\pi^2}
{|\delta_k|}^2{}_{\text{re-entry}}
~=~ \mathcal{A}^2
\left[ 1 + c_0 {\left(\frac{k}{2\pi H}\right)}^{4(1-n)}
\right]
\label{PowerSpectrum}
\end{equation}
where
$c_0 = \frac{\pi^2}{\sigma^2}[1 + \frac{\sigma^2}{3 (1+\sigma)} ]$ and
the $\mathcal{A}^2$ is given by
\begin{equation}
\mathcal{A}^2 ~=~ 
\frac{{(1+\omega_{re})}^2} {(2\pi)^2~C_{\omega}^2}   
~\frac{~ {\sigma}^3~(1+\sigma)~e^{-\sigma} }
{2^{2-\frac{2}{n}}~ {\Gamma(1-\frac{1}{2n})}^4 } ~.
\label{PSAmp}
\end{equation}
The quantity $\sigma$, defined in (\ref{CutoffScale}), is given
by $\sigma ~=~ 4\pi{\left(\frac{2\pi}{3}\right)}^{\frac{2}{3}}
~{\left(\frac{\bar{\rho}}{M_p^4}\right)}^{\frac{1}{3}}$.  For the
super-horizon modes ($\frac{k}{2\pi H} << 1$) the second term in
the expression (\ref{PowerSpectrum}) is negligible compared
to unity for the effective loop quantum cosmology
($-\infty<n<-2$).  Thus it is clear from the expression
(\ref{PowerSpectrum}) that power spectrum of density perturbation
is broadly {\em scale-invariant} as during the inflationary
period the Hubble parameter remains almost constant. The mode
function solutions (\ref{NormalizedMFSolution}) being ambiguity
parameter dependent, expectedly the expression of the power
spectrum (\ref{PowerSpectrum}) depends on ambiguity parameter.
But it should also be noted that this dependence is rather weak. The
ambiguity parameter dependent term in the power spectrum
${2^{-\frac{2}{n}}~ {\Gamma(1-\frac{1}{2n})}^4}$ varies only
between $1$ to $1.3499$ for the range of ambiguity parameter
value ($-\infty<n<-2$).

An important property of the power spectrum (\ref{PowerSpectrum})
is that $\mathcal{A}^2 \sim H^2$ ($\sigma$ being small
$(1+\sigma)~e^{-\sigma} \approx 1$). This behaviour is exactly
similar to the behaviour of the power spectrum in standard
inflationary scenario. This property of the power spectrum will
be very useful in comparison of spectral index between standard
inflationary scenario and effective loop quantum cosmology
scenario.

\subsection{Amplitude of Density Perturbation}

In the section IV, we have shown that the self-consistency of the
framework that we are using requires $\sigma<<1$. This
requirement can be physically understood in the following way.
The effective continuum (classical geometrical) description is an
emergent description in the loop quantum cosmology framework in
which underlying geometry is fundamentally discrete. Naturally
the effective Hamiltonian description which has been used in the
paper has a restricted domain. In \cite{GenBounce}, it has been
shown that the dynamics described by the effective Hamiltonian
respects it own domain of validity provided the permissible
values of $\bar{\rho}$ is chosen to be significantly smaller than
unity when written in Planck units. In other words the
requirement $\sigma << 1$, essentially defines the domain of
validity of the effective Hamiltonian. On contrary, in a purely
classical geometrical description (standard inflationary
scenario) such restriction does not arises as the description
itself a fundamental description of nature within the setup of
general relativity. Thus it is clear from the expression
(\ref{PSAmp}) that in this scenario the amplitude for the power
spectrum of density perturbation is {\em naturally} small. In
other words, the small amplitude for the primordial density
perturbation is a {\em prediction} of the framework of effective
loop quantum cosmology.

It should be noted from the equation (\ref{BardeenFormula}) that
if the equation of state is very close to $-1$ during
inflationary period then the amplitude of the density
perturbation gets a large multiplicative factor at the time of
horizon re-entry.  In the case of loop quantum cosmology induced
inflation the equation of state (\ref{EffEOS}) indeed very close
$-1$ as $a<<1$ during its inflationary period. However in this
scenario, still one can produce small amplitude for the
primordial density perturbation without fine tuning. One of the
reasons behind this is the presence of the small factor
$a^{4(1-n)}$ in the two-point density correlation function
(\ref{TwoPointDC2}).  The presence of this crucial small factor
in the expression of the two-point density correlation function
simply follows from the {\em modified} mode functions of the
scalar field.

Another interesting property of the amplitude (\ref{PSAmp}) is
that it contains an exponential damping term $e^{-\sigma}$.  For
the purpose of this paper the damping term is insignificant as
$\sigma$ is required to be small in this paper. However if one
naively takes the energy scale to be order of Planck scale even
then amplitude of the density perturbation will remain small as the
exponential term becomes significant in that scale. In fact this
was the main motivation of the papers \cite{PaddyII,PaddyI}.

{\em Qualitatively} the smallness of the amplitude for the
primordial density perturbation is readily predicted but to have
{\em quantitative} estimate one needs to choose some value for
the associated energy scale.  Assuming density perturbation is
{\em adiabatic i.e.} it is same as curvature perturbation, one
can relate amplitude of the power spectrum of the density
perturbation to the CMB angular power spectrum as follows
\cite{Riotto:ICTP} 
\begin{equation} \mathcal{A}^2
~=~ ~\left(\frac{3}{2}\right)^2~ \frac{9}{(2\pi^2)}~
\frac{\tilde{l}(\tilde{l}+1)C_{\tilde{l}}^{AD}}{2\pi} ~. 
\label{CMBPS} 
\end{equation}
where $\tilde{l}$ is the multipole number of the angular power
spectrum. We have also assumed that the relevant modes re-enter
horizon during radiation dominated era. The COBE data implies
that $\frac{\tilde{l}(\tilde{l}+1)C_{\tilde{l}}^{AD}}{2\pi}
\simeq 10^{-10}$. Using the expression (\ref{PSAmp}) one can
easily deduce that $\sigma ~\approx~ 5.5\times10^{-3}$(we have assumed
here that at the end of loop quantum cosmology induced inflation,
the radiation domination begins {\em i.e.} $C_{\omega}=4$; value
of the ambiguity parameter $l$ is chosen to be $\frac{1}{2}$).
It follows from the expression of $\sigma$ (\ref{CutoffScale})
that the corresponding energy density is $\bar{\rho} \approx
{(2.0\times10^{-3}~ M_p )}^4 =
\left(2.0\times10^{16}\text{GeV}\right)^4$. The associated 
energy scale comes down slightly if one assume that the
end of loop quantum cosmology induced inflation is followed
by a standard accelerating phase ( For example, if
one takes $C_{\omega}=\frac{2}{3}$ ($\omega=-\frac{1}{2})$ then
the corresponding energy density is $\bar{\rho} \approx{(0.8\times10^{-3}~
M_p )}^4 = \left(0.8\times10^{16}\text{GeV}\right)^4$).

The energy scale required to produce observed amplitude of
density perturbation in the effective loop quantum cosmology is
{\em not} very different from the standard inflationary scenario
where associated energy density is $\bar{\rho} \approx
\left(2.0\times10^{16}\text{GeV}\right)^4$ \cite{Riotto:ICTP}.
Then it is quite important to understand why is it necessary to
{\em fine tune} field theoretical parameters in standard scenario
to produce small amplitude. In this paper we have considered a
massive scalar field as a matter source. The energy density
during standard inflationary period is $\bar{\rho} \approx
\frac{1}{2}~ m_{\phi}^2~\phi^2$. In standard inflationary
scenario to produce sufficient amount of expansion (to solve
horizon problem and others) one needs to choose the values of
field to be $\phi^2 \approx 10 M_p^2$ \cite{Riotto:ICTP}. This in
turns forces one to tune the mass parameter to be $m_{\phi}
\approx 10^{-6} M_p$, in order to produce small amplitude for
primordial density perturbation. These fine tunings of field
strength and mass term are not only severe but also extremely
difficult to sustain under standard quantum field theory.
Specifically, sustaining such low mass parameter from loop
corrections often requires new ingredients \cite{Liddle:CILSS}.

In other words, in the standard inflationary scenario to get
correct amplitude of density perturbation, the required fine
tunings are directly related with the method by which inflation is
realized namely the imposition of {\em slow-roll} condition.  One
the other hand, in loop quantum cosmology the inflationary phase
is realized generically \cite{GenericInflation} and {\em not} by
imposing slow-roll condition. There is a physical understanding
of it from the fundamental point of view. The famous singularity
theorems tell us that in the cosmological set-up one cannot avoid
initial singularity if the matter contents always satisfy the
strong energy condition (equation of state $\omega \ge -
\frac{1}{3}$). Naturally to avoid initial singularity the quantum
gravity effects must drive the matter contents to effectively
violate the strong energy condition, atleast for small enough
volume. It has been shown that in loop quantum cosmology one
generically avoids singularity at the quantum level \cite{Sing}
as well as the effective classical level
\cite{Singh:Bounce,Vereshchagin,GenBounce}. Using Raychaudhuri
equation it is then easy to see that the violation of strong
energy conditions i.e. $\omega < -\frac{1}{3}$ immediately
implies an accelerating phase. In particular in
\cite{GenericInflation}, it has been proved that for {\em any}
positive definite scalar potential loop quantum cosmology induced
accelerating phase undergoes near exponential expansion.  Thus,
in loop quantum cosmology the realization of an early
accelerating phase is intimately related with the removal of
initial singularity. In this scenario, the inflation is
essentially driven by the non-perturbative modification of the
scalar field dynamics namely via the spectrum of the inverse
scale factor operator. This inflationary phase begins as soon as
the system gets into the small volume regime (non-perturbative
domain). Therefore, one does {\em not} require to fine tune the
field strength to sufficiently uphill like as done in the
standard scenario. Consequently, one does not require to fine
tune field theoretical parameters to produce small amplitude of
density perturbation. Rather, as it is shown in this paper, the
small amplitude is a prediction of the framework that has been
used in the calculation.

Nevertheless one can impose self-consistency requirement on the
mass parameter in this calculation. Let us recall that in
simplifying mode function equation (\ref{ModifiedMFEquation}) we
have neglected the mass term $\frac{m_{\phi}}{H}$ compared to the
term $\frac{k}{H a}$.  Since we are interested in calculating
power spectrum at the time of horizon crossing {\em i.e.}
$\frac{k}{H a}=2\pi$ then to be self-consistent we must require
that $m_{\phi}<2\pi H$. This in turns implies that the maximum
value of the mass parameter to be $m_{\phi} ~\sim~
\frac{1}{27}{~(\bar{\rho})}^{\frac{1}{4}}$. We may mention here
that in effective loop quantum cosmology $\bar{\rho}$ in fact is
the maximum energy density {\em i.e.}
${(\bar{\rho})}^{\frac{1}{4}}$ is precisely the cut-off scale. It
is worth emphasizing that the restriction on mass parameter here
is a self-consistency requirement of the calculation as solving
the modified mode function equation (\ref{ModifiedMFEquation})
including the mass term turns out to be {\em not} so easy a task.

\subsection{Spectral Index}

So far in the calculation we have assumed that during
inflationary period energy density is strictly constant. But this
was rather an approximation to simplify the calculation. We can
in fact compute the effect of small variation of energy density.
The small variation of Hubble parameter leads to a small
deviation from the {\em scale-invariant} power spectrum. The
scale dependent property of the power spectrum is conveniently
described in terms of {\em spectral index}. From the conservation
equation it follows that $\frac{d~ \text{ln} \rho}{d~ \text{ln}
a} = -3 (1+\omega^{\text{eff}}) = -C_{\omega}(n+2) a^{2(1-n)}$.
Using this relation one can compute the spectral index at horizon
crossing 
\begin{eqnarray}
n_s - 1 ~&:=&~ \frac{d~ \text{ln} \mathcal{P}_{\delta}(k)}{d~ \text{ln k}}
\nonumber \\
~&\approx&~ C_{\omega}(-n-2) {\left(\frac{k}{2\pi H}\right)}^{2(1-n)}
~+~ 4 c_0 (1-n) {\left(\frac{k}{2\pi H}\right)}^{4(1-n)} ~.
\label{SpectralIndex}
\end{eqnarray}
We may note here that the spectral index $n_s$ is {\em extremely}
close to unity and the difference $(n_s-1)$ depends non-trivially
on the ambiguity parameter. But the most important property of
the spectral index (\ref{SpectralIndex}) is $(n_s-1)>0$ for all
allowed values of the ambiguity parameter ($0<l<1$ {\em i.e.}
$-\infty<n<-2$). This is in complete contrast to the standard
single-field inflationary scenario. We have mentioned earlier
that the power spectrum for both the scenarios varies as $H^2$.
For single-field standard inflationary scenario the leading
contribution to the spectral index deviation comes from the
variation of Hubble parameter during inflationary period. So for
the standard scenario the spectral index is given by
\begin{equation}
n_s - 1 ~:=~ \frac{d~ \text{ln} \mathcal{P}_{\delta}(k)}{d~ \text{ln k}}
~\approx~ -6~\epsilon ~,
\label{SSpectralIndex}
\end{equation}
where $\epsilon=\frac{1}{16\pi G}
{\left(\frac{V'(\phi)}{V(\phi)}\right)}^2 = 4\pi G \frac{{\dot
\phi}^2}{H^2}$ is the {\em slow roll} parameter of the standard
inflationary scenario. In fact due to the time variation of
$\epsilon$ there will be additional contributions in
(\ref{SSpectralIndex}). But those will be {\em sub-leading} for
{\em single-field} inflationary scenario. Thus for single-field
standard inflationary scenario the spectral index satisfies
$(n_s-1)<0$.

It is worth pointing out that in the standard method (indirect
method) of obtaining power spectrum of the density perturbation,
one first computes the power spectrum of the field fluctuation.
Using the expression (\ref{ApproxMF}), one can compute the
power spectrum of the {\em field} fluctuation at the horizon exit,
given by
\begin{equation}
\mathcal{P}_{\phi}(k)  := \frac{k^3}{2\pi^2} {|f_k|}^2
= \left(\frac{1}{2\pi}\right)
\frac{|n+2|}{3} ~ 
\frac{n^2~ 2^{\frac{1}{n}}} { {\Gamma(-\frac{1}{2n})}^2 }~ H^2, 
\label{LQCPPofFF}
\end{equation}
which is {\em scale invariant}. In the standard inflationary
scenario, this would have lead to a scale invariant power
spectrum of {\em density} perturbation. However, we have already
mentioned that this is not straightforward in the effective loop
quantum cosmology. This is due to the fact that it is not yet
well settled; how to obtain a general effective action,
consequently an effective stress-energy tensor from a canonical
quantum theory of gravity. This prevents one to compute the power
spectrum of density perturbation using standard method, as one
needs to know the expression of the general effective
stress-energy tensor, to relate the power spectrums of field
fluctuation to the density perturbation. Nevertheless, one may
naively assume that the power spectrum of field fluctuation and
that of density perturbation will have similar relation. In that
case, the computed power spectrum of density perturbation using
{\em direct} and {\em indirect} method would differ from each
other by an extremely weak $k$ dependence.  This difference would
then be similar in nature to the results of \cite{PaddyII} in the
context of the standard inflationary scenario.  However this
difference would at most affect the quantitative nature (that too
by a vanishingly small amount) but {\em not} the qualitative
nature of the spectral index (\ref{SpectralIndex}) whatsoever.

Thus it is clear that for {\em effective loop quantum cosmology}
induced inflationary scenario the spectral index has a
qualitatively distinct feature compared to that of {\em
single-field} driven {\em standard} inflationary scenario. In the
next sub-section we discuss its observational consequences.

\subsection{Observational Implications}

The power spectrum of density perturbation
generated during inflationary era is {\em not} directly
observable. Rather the observed power spectrum corresponds to
the density perturbation at the time of {\em re-entry} in the
post-inflationary period. At the time of re-entry larger
wavelength ($2\pi k^{-1}$) enters the horizon at later time compared
to the smaller wavelength. It is clear from the expression
(\ref{BardeenFormula}) that if there is a change in the equation
of state of the universe during re-entry then there will be an
additional modification of the power spectrum. Since we are
interested in estimating the original power spectrum generated
during inflationary era then its quite important to avoid
additional modification of the power spectrum coming from 
other possible sources.

The observed anisotropy in the CMB sky corresponds to the density
perturbation on the {\em last scattering surface}. The last
scattering surface broadly demarcate the end of radiation
domination era to the beginning of matter domination era.
Naturally during this period ($1+\omega$) changes from
$\frac{4}{3}$ to $1$. While deriving the expression
(\ref{SpectralIndex}) of the spectral index we have assumed
constant equation of state during re-entry. Thus for the purpose
of comparison with observations, one must consider only those modes
for which the equation of state was almost constant during 
re-entry. On last scattering surface they will corresponds to
the modes which are well inside the horizon at the time of {\em
decoupling}. Being smaller in wavelength these mode will subtend
smaller angle in present day sky. Naturally these mode will
corresponds to the higher multi-pole number. Also if one
considers sufficiently narrow bands in these part of
spectrum then one can avoid additional modification coming from
the sub-horizon evolution of density perturbation in the
period between their {\em re-entry} and the decoupling. 

To infer the property of primordial density perturbation from the
observed angular power spectrum of CMB, one needs to know the
evolution of the universe for the period between the decoupling
and the present day universe. Since major fraction of today's
energy density is believed to be coming from mysterious {\em dark
matter} and {\em dark energy} then it is quite obvious that there
will be a considerable influence of them on the inferred
primordial power spectrum. The current observational estimate of
spectral index based on WMAP+SDSS data is $n_s=0.98 \pm 0.02$
\cite{Tegmark:Inflation,Tegmark:CP,Seljak}. This estimate is
based on the entire part of the observed angular power spectrum
nevertheless this agrees (rather marginally) with the expression
of the spectral index (\ref{SpectralIndex}) which is strictly
valid only for the part of the spectrum in the higher multipole
region. For this purpose, it may be more convenient to {\em
reconstruct} the primordial power spectrum using observed CMB
angular power spectrum (for example as done in
\cite{Bridle,Mukherjee,Hannestad,Tarun}) and then consider the
higher wavenumber part of the spectrum.

\section{Discussions}

In summary, we have computed the power spectrum of density
perturbation generated during loop quantum cosmology induced
inflationary phase. The resulting power spectrum is broadly {\em
scale-invariant}. Further it has been shown that the small
amplitude for primordial density perturbation is a natural {\em
prediction} of the framework of effective loop quantum cosmology.
Unlike standard inflationary scenario, here one does not require
to {\em fine tune} field theoretical parameters to produce
observed small amplitude for density perturbation. The resulting
power spectrum also has a qualitatively distinct feature
compared to the standard single-field inflationary scenario.
The spectral index in the effective loop quantum cosmology
scenario satisfies $(n_s-1)>0$ whereas for the standard
inflationary scenario it satisfies $(n_s-1)<0$.

Naturally, the spectral index of power spectrum for density
perturbation generated during the loop quantum cosmology induced
inflation and the standard inflation differs from each other in a
non-trivial and non-overlapping way. This is a consequence of the
fact that the during loop quantum cosmology induced inflation the
Hubble horizon {\em shrinks} marginally whereas in the standard
inflationary scenario the Hubble horizon {\em expands}. This
feature leads to the power spectrum for the corresponding density
perturbation to be tilted in opposite directions to each other.
We have argued that this feature is a {\em generic} property of
the corresponding scenario and not a property of some particular
model. We have also pointed out the part of the observed CMB
angular power spectrum that may be better suited for testing this
particular feature observationally, namely the part corresponding
to the higher multipole numbers of the CMB angular power spectrum.

It is worth pointing out that the calculation techniques used
here within the adopted framework are analytic and the
approximations used here are mostly justified. Nevertheless one
should keep it in mind that this calculation itself  should {\em
not} be considered as the {\em first principle} calculation of
density perturbation within loop quantum cosmology. Rather this
calculation is based on {\em effective} loop quantum cosmology.
Here we have considered the non-perturbative modification of
kinetic term of the scalar matter Hamiltonian. In a first
principle calculation (using inhomogeneous model), one may
naively expect to get corrections also in the gradient term of
the matter Hamiltonian. This modification should depend on some
ambiguity parameter similar to that of $l$ here.  In this paper
it is shown that the ambiguity parameter $l$ dependence of the
amplitude of the power spectrum is very weak. So the effect of
such possible modifications on the amplitude of the power
spectrum is expected to be rather small. The effect on spectral
index deviation due to this modification is also expected to be
small as it is determined mainly by the background evolution
itself. Thus it is very likely that the calculation presented
here should be a good approximation of what is expected from a
first principle calculation in the energy scale concerned.

We have used the direct method to compute the power spectrum of
density perturbation.  This method uses the same techniques which
are used in the standard quantum field theory. So one needs to
have some kind of ultra-violet divergence regularization
prescription.  To regularize the ultra-violet divergence we have
used the method outlined and explicitly shown in
\cite{PaddyII,PaddyI}. This method relies on the assertion that a
proper theory of quantum gravity should contain a {\em
zero-point} proper length. In the effective loop quantum
cosmology such length scale {\em is} naturally available. In this
method, regularization is essentially carried out by adding a
{\em zero-point} proper length to the standard definition of
proper length. In this paper the procedure was notationally
simplified by using the notion of {\em effective coordinate
length}.  Nevertheless regularization procedure was carried out
{\em by hand}. But this was expected as the calculations here
were done using standard quantum field theory. On the other hand
one would expect that in a {\em first principle} calculation
these regulator should come {\em built-in} as it has been argued
in the context of full theory \cite{Thiemann:QSDV}. The crucial
result of the paper namely the properties of the spectral index
are insensitive to the regularization method as it is dominantly
determined by the nature of the effective equation of state
during its inflationary phase.

In standard inflationary scenario one is also interested in
computing power spectrum for tensor mode of the metric
fluctuations {\em i.e.} gravitational wave. But as we have
mentioned that the technology required to deal with inhomogeneity
at fundamental level in loop quantum cosmology is not available
yet. In loop quantum gravity approach {\em geometry} is quantized
in non-perturbative way. Thus it is not very easy to `guess' the
structure of the quantum fluctuation of geometry until one does
explicit calculation within the framework.  Nevertheless one
would naively expect that the power spectrum for tensor mode
perturbation should be similar to the standard scenario as the
structure of the effective gravity sector Hamiltonian is similar
to the classical Hamiltonian in the relevant length scale, apart
from the fact that the relevant energy scale during corresponding
inflationary period is also similar.

Before we discuss the implications of possible outcomes of
mentioned test let us have a comparative study of standard
inflationary scenario and loop quantum cosmology induced
inflationary scenario. In order to have a {\em successful}
inflation in the standard scenario, generally one requires
multi-level of fine tuning of field parameters. In other words
one faces several kind of {\em naturalness} problems to achieve
a successful inflation.

{\em The first} one is to {\em start inflation}. In standard
inflationary scenario it is needed to choose initial field
velocity to be sufficiently small so that the equation of state
$\frac{\frac{1}{2} {\dot \phi}^2 -V(\phi)} {\frac{1}{2} {\dot
\phi}^2 +V(\phi)} \approx -1$.  {\em The second} one is to {\em
sustain inflation}. In standard scenario one requires to choose
the field potential to be {\em flat} enough so that field does
not gain momentum quickly. {\em The third} one is to {\em
generate sufficient expansion} (to solve horizon problem and
others). To achieve this in standard scenario, one requires to
choose the initial field configuration sufficiently {\em uphill}
in the potential. In other words, one requires to fine tune
initial field configuration. {\em The fourth} one is to {\em end
inflation}. In many cases this requires sort of {\em potential
engineering} to have a {\em long flat plateau} and then a {\em
fast fall-off} in the potential profile. {\em The fifth} one is
to produce {\em small amplitude} for primordial density
perturbation. To produce observed small amplitude of density
perturbation one needs to fine tune parameters of the field potential.
This fine tuning is basically required to {\em compensate} the `third'
fine tuning.

On the other hand, to achieve the first, second and fourth
requirements in loop quantum cosmology induced inflationary
scenario one {\em does} not require to fine tune the parameters.
These requirements are naturally achieved as they simply follow
from the spectrum of the inverse scale factor operator. The fifth
requirement {\em i.e.} small amplitude, as shown in this paper,
is a natural prediction of effective loop quantum cosmology.  The
situation regarding the third problem also gets improved
significantly. In the loop quantum cosmology, the generated
amount of expansion is controlled by the ambiguity parameter $j$.
Clearly to produce sufficiently large expansion, using loop
quantum cosmology {\em alone}, one will require to choose the
value of $j$ to be large. Thus it is very likely that only the
initial part of the inflation was driven by loop quantum
cosmology modification. It has been argued in \cite{Bojowald:BPI,
Tsujikawa, Bojowald:QMC} that the loop quantum cosmology induced
inflationary phase can lead to a secondary standard inflationary
phase. This follows from the fact that the in-built inflationary
period of loop quantum cosmology can produce favourable initial
conditions for an additional standard inflationary phase. In
\cite{Tsujikawa}, the authors have also studied the possible
effects of the above mechanism on CMB angular power spectrum
generated during the standard inflationary phase that follows the
LQC induced inflationary phase and shown that it can lead to
suppression of power in the low CMB multipoles.  Since the
observed part of CMB angular power spectrum generally corresponds
to early period of inflation then it may well be the situation
where the observed part of the CMB angular power spectrum
corresponds to the loop quantum cosmology driven inflationary
period.

It is worthwhile to emphasize that high amount of expansion in this
scenario is required {\em not} to solve horizon problem (being
non-singular this model avoids horizon problem
\cite{GenericInflation}) rather to avoid a different kind of
problem. We have seen that the `initial size' of universe was
typically order of Planck units and the corresponding energy
scale was also typically order of Planck units. During
relativistic particle (radiation) dominated era energy scale
falls of typically with inverse power of the associated length
scale.  It is then difficult to understand why the universe is so
large ($\sim 10^{60} L_p$) today but still it has relatively very
high energy scale ($\sim 10^{-30} M_p$). During inflationary
period, on the other hand, the energy scale remains almost
constant whereas the length scale grows almost exponentially with
coordinate time. It is now clear that we can avoid this
discrepancy between energy scale and the length scale of the
universe provided there existed an inflationary period with
sufficiently long duration in early universe.

Now if the observed power spectrum turns out to be not in
agreement with the computed power spectrum, then one should
conclude that the phase of inflation corresponding to the
observed window couldn't possibly be driven by loop quantum
cosmology modification. It may then restrict the allowed choices
for the ambiguity parameter $j$. Consequently it will be an
important issue to understand within the framework of {\em
isotropic} loop quantum cosmology with {\em minimally coupled}
scalar matter field, why the observed universe today is so large
but still it has sufficiently high energy scale.

\begin{acknowledgments}
I thank Ghanashyam Date and Martin Bojowald for helpful
discussions and critical comments on the manuscript. I thank
Martin Bojowald for a cordial invitation to visit
Max-Planck-Institut f\"ur Gravitationsphysik where a major part
of this work was completed. I also thank the anonymous referee
whose constructive as well as instructive criticisms has lead to
significant improvement of the paper and illumination of several
conceptual issues.
\end{acknowledgments}


\end{document}